\documentclass[prd,twocolumn,showpacs]{revtex4}
\usepackage{hyperref,amssymb,amsmath,mathrsfs,bm,graphicx}
\usepackage{enumitem}
\usepackage{color}
\begin{document}
\title{Newtonian polytropes for anisotropic matter: General framework and applications}
\author{L. Herrera}
\email{laherrera@cantv.net.ve}
\affiliation{Departamento de F\'\i sica Te\' orica e Historia de la Ciencia,
Universidad del Pa\'\i s Vasco, Bilbao 48940, Spain}
\author{W. Barreto}
\email{wbarreto@ula.ve}
\affiliation{Centro de F\'\i sica Fundamental, Facultad de Ciencias,
Universidad de Los Andes, M\'erida 5101, Venezuela}
\begin{abstract}
We set up the general formalism to model polytropic Newtonian stars with anisotropic pressure. We obtain the corresponding Lane-Emden equation.  A heuristic model based on an ansatz to obtain anisotropic matter solutions  from known solutions for isotropic matter is adopted to illustrate the effects of the pressure anisotropy on the structure of the star.  In particular, we calculate the Chandrasekhar mass for a white dwarf. It is clearly displayed how the Chandrasekhar mass limit changes depending on the anisotropy. Prospective astrophysical applications of the proposed approach are discussed.
\end{abstract}
\date{\today}
\maketitle

\section{Introduction}
In the context of Newtonian gravity,   polytropic equations of state are particularly useful to describe a great variety of situations (see Refs. \cite{1, 2, 3, 8, 9,10} and references therein),  their great success stemming mainly  from  the simplicity of the equation of state and the ensuing main equation (Lane-Emden).  Polytropes in the context of general relativity have been considered in  Refs. \cite{4a,4b,4c,5,6,7,11,12,13} (and references therein). However, in this work, we restrict the analysis to Newtonian polytropes.

The theory of polytropes is based on the polytropic equation of state
\begin{equation}P=K\rho^{\gamma}=K\rho^{1+1/n} ,\label{Pol}\end{equation}
where $P$ and $\rho$ denote the isotropic pressure and the  mass (baryonic) density,  respectively. Constants $K$, $\gamma$, and $n$ are usually called  the polytropic constant, polytropic exponent, and polytropic index, respectively.

The polytropic equation of state may be used to model two very different types of situations, namely:
\begin{enumerate}[label=(\roman*)]
\item When the polytropic constant $K$ is fixed and can be calculated from natural constants. This is the case of a completely degenerate gas in the nonrelativistic ($\gamma=5/3; n=3/2$) and relativistic limit ($\gamma=4/3; n=3$). Polytropes of this kind are particularly useful to model compact objects such as withe dwarfs (WDs), and they lead in a rather simple way to the 
Chandrasekhar mass limit.
\item When $K$ is a free parameter, as, for example, in the case of isothermal ideal gas, or in a completely convective star.
Models related to isothermal ideal gas are relevant in the so-called Sch\"onberg--Chandrasekhar limit (see Ref. \cite{3} for details).
\end{enumerate}

Our motivation to extend  polytropic stellar models to cases in which the pressure anisotropy is allowed is based on the fact that the local anisotropy of pressure may be caused by a large variety of physical phenomena of the kind  we expect in compact objects (see Ref. \cite{14} for an extensive discussion on this point). 

Indeed, the study  of anisotropic (principal stresses unequal) spherically symmetric fluids  has a long and  venerable story.
It started with Jeans \cite{jeans}, who studied the anisotropy produced by anisotropic velocity distributions in galaxies.
The first mention of local anisotropy of pressure in spherically symmetric  selfgravitating fluids may be found in the seminal paper by Lemaitre \cite{lemaitre}. In page 63 of that paper, Lemaitre realizes that the stringent limit in the compactness of a homogeneous relativistic sphere is related to the isotropy of pressure, and therefore  he proposes to relax that condition. He considers the ``limiting" case in which the radial pressure vanishes, but the tangential does not.
However, the interest in this subject started to grow exponentially after the pioneering work of Bowers and Liang \cite{bl}.  For recent references on this subject, see Refs.  \cite{hdmost04, hls89, hmo02, hod08, hsw08} (and references therein). 
{An alternative approach to anisotropy comes from kinetic theory
using the spherically symmetric Einstein-Vlasov equations, which admits a very rich class of static solutions, none of them isotropic (Ref. \cite{andreasson} (and
references therein). The advantages or disadvantages of either approach are related to the specific problem under consideration. As we shall see below, our method 
links our models continually with the isotropic case (see Sec. IV),
thereby allowing us to bring out the influence of anisotropy on the structure
of the object. Evidently, both approaches should give the same physical
results.}

Among all possible sources of anisotropy,  there is one  particularly related to our endeavor in this manuscript, namely the intense magnetic field observed in compact objects such as white dwarfs,  neutron stars, or magnetized strange quark stars  (see, for example, Refs. \cite{15, 16, 17, 18, 19} and references therein).

Indeed, it is a well-established fact  that  a magnetic field acting on a Fermi gas produces pressure anisotropy (see  Refs. \cite{22, 23, 24, 25, 26} and references therein). In some way, the magnetic field can be addressed as a fluid anisotropy. 

Particularly appealing is the fact that  magnetic fields may severely affect the Chandrasekhar mass limit of a white dwarf  \cite{27, 20, 21}.

For all the reasons above, we intend in this paper to develop the general formalism to describe polytropes in the presence of pressure anisotropy. 

For the  sake of completeness, we shall first  review very briefly the theory of polytropes for a perfect (isotropic) fluid. Next, we shall display the general formalism  for anisotropic fluids. In  order to bring out the effects of anisotropy on the structure of the star, we shall further assume an ansatz  allowing us to calculate the influence of anisotropy on the Chandraskhar mass limit.
Finally, we shall conclude with some  possible applications and  unanswered issues.

\section{The polytrope for fluids with isotropic pressure}

 Polytropes are assumed to be in hydrostatic equilibrium (for deviations from this condition see Refs. \cite{8} and \cite{11});
 therefore, the two starting equations are the equation of hydrostatic equilibrium,

\begin{equation}\frac{dP}{dr}=-\frac{d\Phi}{dr} \rho,
\end{equation}
and the Poisson equation (in spherical coordinates),

\begin{equation}\frac{1}{r^2} \frac{d}{dr}(r^2\frac{d\Phi}{dr})=4\pi G \rho,
\label{Poisson}
\end{equation}
with $\Phi$   and $G$ denoting the Newtonian gravitational potential and the gravitational constant respectively.

Combining the two equations above with Eq. (\ref{Pol}), one obtains after some simple calculations the well-known Lane-Emden equation (for $\gamma\neq 1$):
\begin{equation}\frac{d^2\omega}{dz^2}+\frac{2}{z}\frac{d\omega}{dz}+\omega^n=0,\end{equation}
with
\begin{equation}r=\frac{z}{A},\label{ere1} \end{equation}
\begin{equation}A^2=\frac{4 \pi G \rho_{c}^{(n-1)/n}}{K (n+1)},\label{A1} \end{equation}
\begin{equation}\omega^n=\frac{\rho}{\rho_{c}}\label{psi1}, \end{equation}
where the subscript $c$ indicates that the quantity is evaluated at the center, and the following boundary conditions apply: 
$$\frac{d\omega}{dz}(z=0)=0;\,\,\,  \omega(z=0)=1.$$ 
The boundary surface of the sphere is defined by  $z=z_{n}$, such that $\omega(z_{n})=0$. As is well known, bounded configurations exist only for $n<5$, and analytical solutions may be found for $n=0, 1$ and $5$.

In the case $\gamma=1$, which corresponds to an isothermal ideal gas, the  ensuing Lane-Emden equation reads
\begin{equation}
\frac{d^2\omega}{dz^2}+\frac{2}{z}\frac{d\omega}{dz}=e^{-\omega},\end{equation}
where now
\begin{equation}r=\frac{z}{A},\label{ere} \end{equation}
\begin{equation}A^2=\frac{4 \pi G \rho_{c}}{K},\label{A} \end{equation}
\begin{equation}\omega=\frac{\Phi}{K}\label{psi} \end{equation}
 and the following boundary conditions apply: 
 $$\frac{d\omega}{dz}(z=0)=0; \,\,\, \omega(z=0)=0.$$ 

We shall  next generalize the  scheme above  to the case when the pressure is no longer isotropic.

\section{The polytrope for anisotropic fluids}
If we allow the principal stresses to be unequal, then  the hydrostatic equilibrium  equation reads
\begin{equation}
\frac{dP_r}{dr}=-\frac{d\Phi}{dr}\rho + \frac{2}{r}(P_{\bot}-P_r),
\label{hse}
\end{equation}
where $P_r$ and $P_{\bot}$ denote the radial and tangential pressures, respectively. This is the Newtonian limit  of  the generalized Tolman-Opphenheimer-Volkoff equation for anisotropic matter. We recall  that the Tolman-Opphenheimer-Volkoff equation is a consequence of Einstein equations and Bianchi identities, or equivalently, it comes directly from the vanishing of the covariant divergence of the energy momentum tensor. If spherical symmetry is assumed, then necessarily the nonradial stresses are equal: $P_\theta=P_\phi=P_\bot$,
   the only freedom being in this case that $P_r\neq P_\bot$.
   Indeed, spherical symmetry supposes 
   enough freedom to rotate Cartesian axes in order to guarantee 
   $P_x=P_y=P_\bot$ and $P_z=P_r \neq P_\bot$. 
   Of course, if one does not assume spherical symmetry, then in
   principle all three main stresses may be different.

For  the Poisson equation,  of course, we get the same expression [Ref. \ref{Poisson}], as
both are under spherical symmetry.

We shall next assume a  polytropic equation [Eq. (\ref{Pol})] for the radial pressure $P_r$.
Then, using Eqs. (\ref{hse}) and (\ref{Poisson}), we may write
\begin{equation}
\frac{d\Phi}{dr}=-\gamma K \rho^{\gamma-2}\frac{d\rho}{dr}+\frac{2}{r}\frac{\Delta}{\rho},\label{2}
\end{equation}
where $\Delta\equiv P_{\bot}-P_r$. 

For the case $\gamma\neq 1$, we can formally integrate the equation above between any interior $r$ and the surface of radius $r=r_{\Sigma}=constant$, which gives us
\begin{equation}
\Phi=F(r) - K(n+1)\rho^{1/n},
\end{equation}
which can be written as
\begin{equation}
\rho=\left[\frac{(\Phi-F)}{-K(n+1)}\right]^n,
\end{equation}
where
\begin{equation}
F(r)=2\int^r_{r_\Sigma} \frac{\Delta}{r\rho}dr.\label{Fdef}
\end{equation}
Introducing the variables
\begin{equation}
z\equiv \alpha r,
\vspace{0.2cm}
\end{equation}
\begin{equation}
\alpha^2\equiv\frac{4\pi G}{(n+1)^nK^n}[-(\Phi_c-F_c)]^{n-1}
\end{equation}
and
\begin{equation}
w\equiv\left( \frac{\rho}{\rho_c} \right)^{1/n}=\frac{\Phi-F}{\Phi_c-F_c},
\vspace{0.2cm}
\end{equation}
where as before, the subscript $c$ indicates that the quantity is evaluated at $r=0$, the  extended Lane-Emden equation can be written as
\begin{equation}
\frac{d^2X}{dz^2}+\frac{2}{z}\frac{dX}{dz}+(X-Y)^n=0,
\label{le1}
\end{equation}
where 
\begin{equation}
X=w+Y
\end{equation}
and
\begin{equation}
Y=\frac{F}{\Phi_c-F_c}.
\end{equation}

For the isothermal case which corresponds to $\gamma=1$,
 Eq. (\ref{2}) becomes
\begin{equation}
\frac{d\Phi}{dr}=-K\rho^{-1}\frac{d\rho}{dr}+\frac{2 \Delta}{r\rho},
\label{iso1}
\end{equation}
which after integration yields
\begin{equation}
\rho=\rho_{c}e^{(F-\Phi)/K},
\label{2iso}
\end{equation}
where the potential was set to zero at $r=0$ and $F$ is defined by Eq. (\ref{Fdef})
(with the inferior limit set to $r=0$, of course).

Then the corresponding Lane-Emden equation becomes
\begin{equation}
\frac{d^2X}{dz^2}+\frac{2}{z}\frac{dX}{dz}=e^{-w},
\label{iso4}
\end{equation}
with 
\begin{equation}z=\alpha r,\end{equation}
\begin{equation}\alpha^2=\frac{4\pi G\rho_c}{K},\end{equation}
\begin{equation}X=\frac{F}{K}+w=\frac{\Phi}{K}.\end{equation}

It is obvious that in order to proceed further with the modeling of the compact object [i.e., in order to integrate Eqs. (\ref{le1}) or (\ref{iso4})], we need to prescribe the specific anisotropy of the problem ($\Delta$). Such information, of course, depends on the specific physical problem under consideration. Here we shall not follow that direction; instead, we shall assume an ansatz already used in the modeling of relativistic anisotropic  stars \cite{chew81,chew82}, whose main virtue (besides its simplicity) is the fact that the obtained models are continuously connected with the isotropic case.

\begin{figure}
\includegraphics[width=2.in,height=3.in,angle=0]{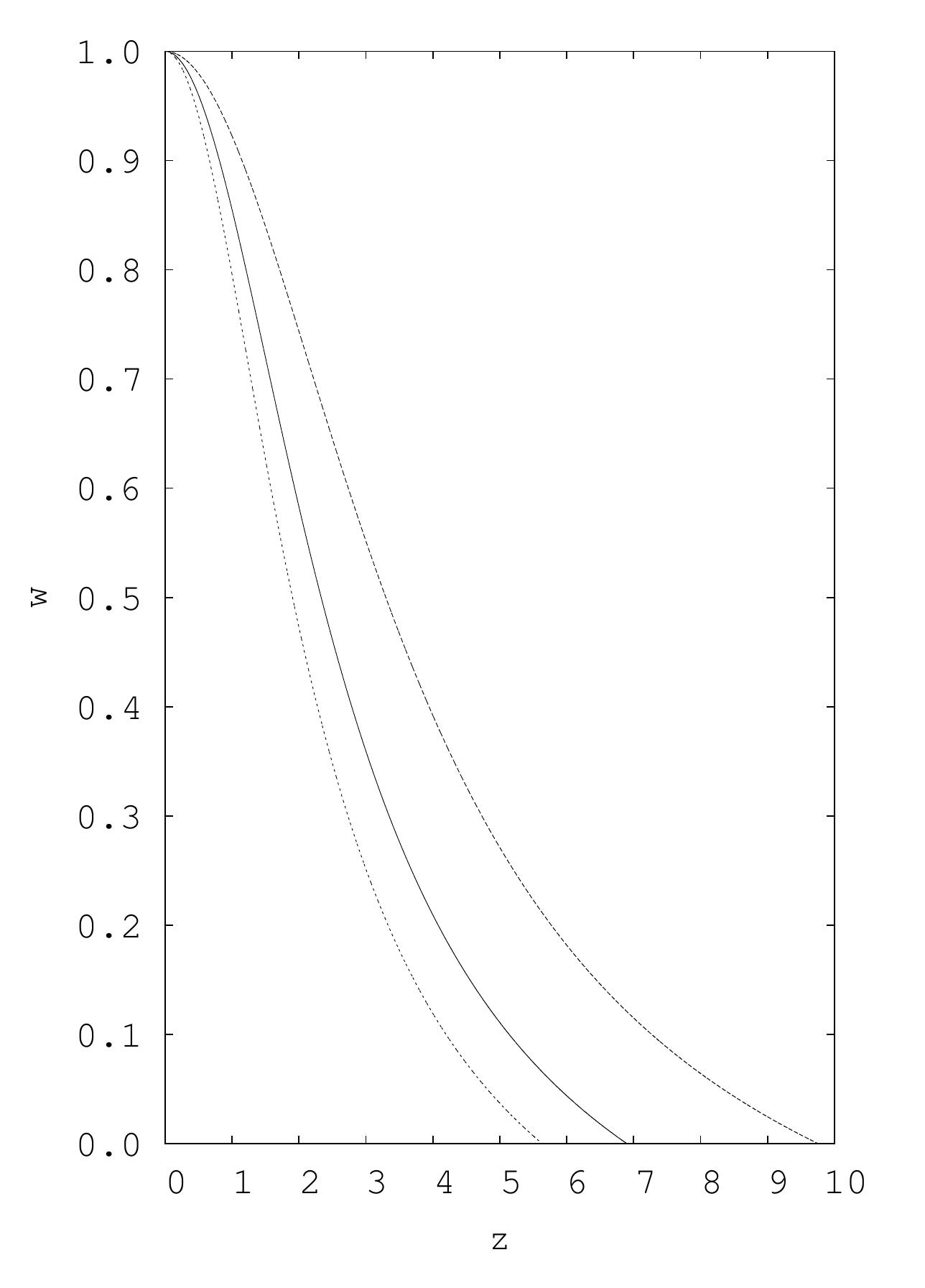}
\caption{$w$ as a function of $z$ for $n=3$ (WD) and $h=0.5$ (dashed line, upper); $h=1.0$ (solid line); $h=1.5$ (short-dashed line, lower).}
\label{fig:wd}
\end{figure}
\begin{figure}
\includegraphics[width=2.8in,height=2.5in,angle=0]{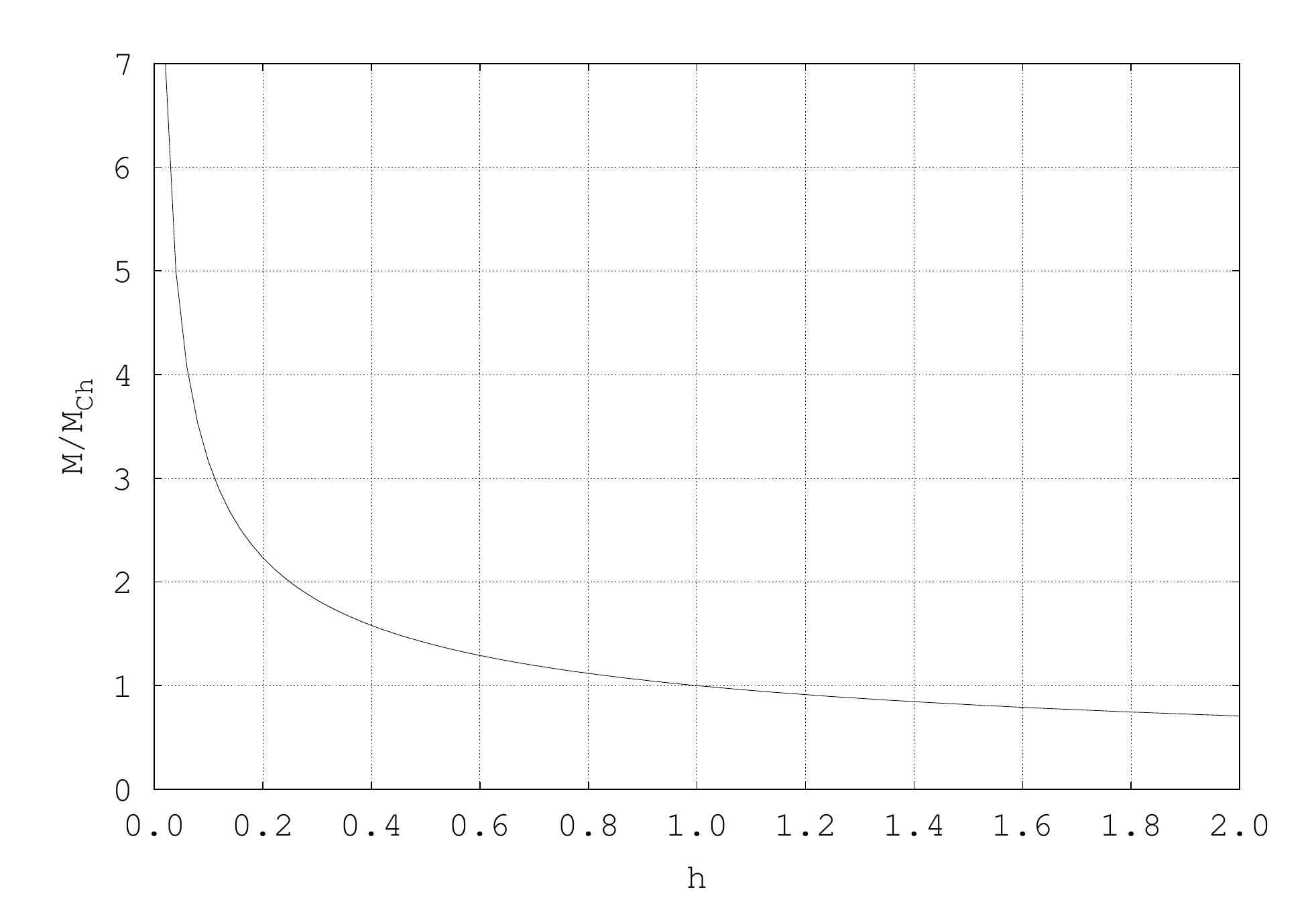}
\caption{Mass ratio $M/M_{Ch}$ as a function of $h$.}
\label{fig:mass}
\end{figure}
\begin{figure}
\includegraphics[width=2.in,height=3.in,angle=0]{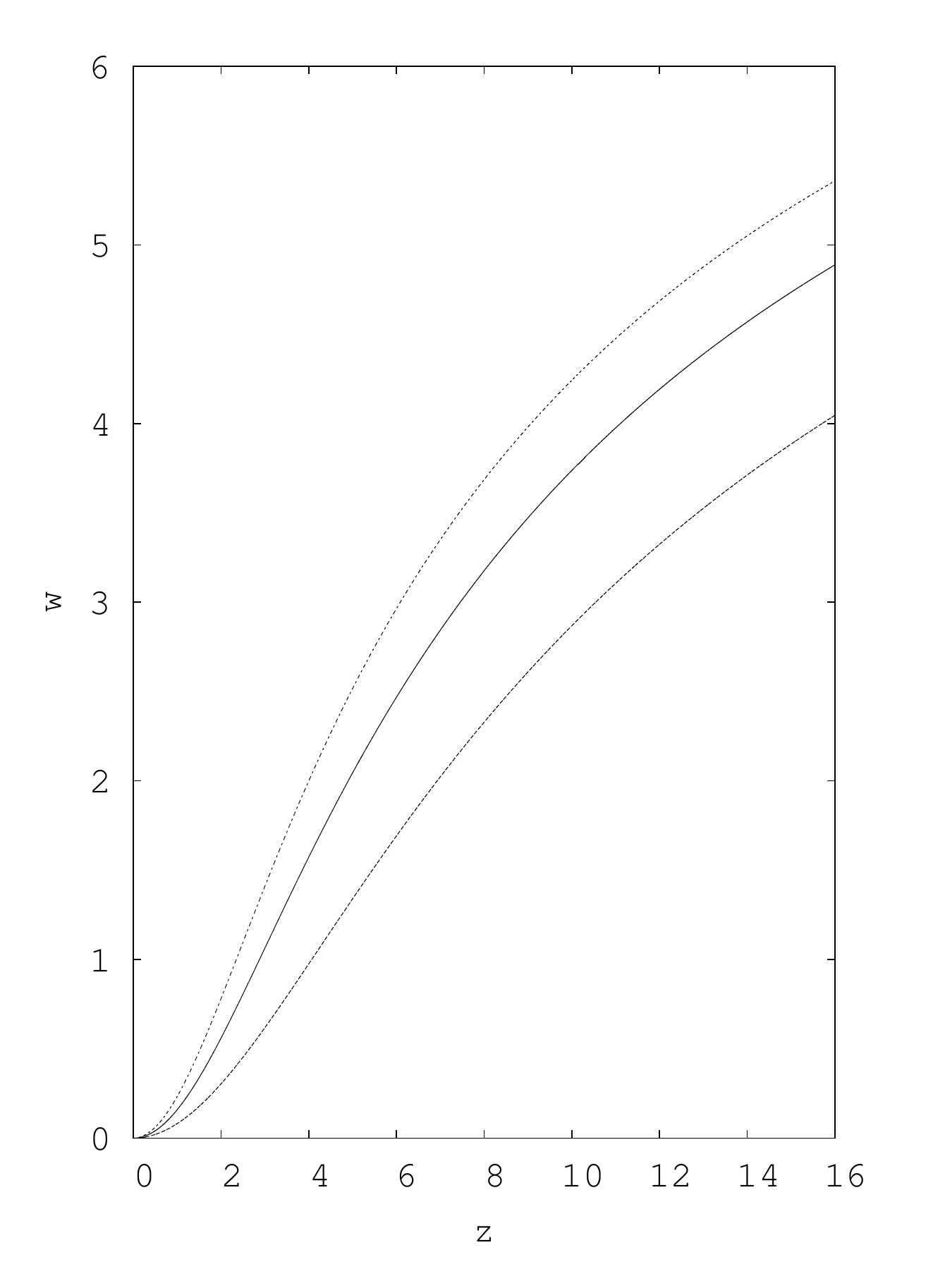}
\caption{$w$ as a function of $z$ for $\gamma=1$ (isothermal gas) and $h=0.5$ (dashed line, lower); $h=1.0$ (solid line); $h=1.5$ (dot-dashed line, upper).}
\label{fig:isothermic}
\end{figure}
\section{Modeling anisotropic polytropes}
In order to obtain specific models, we shall  here adopt the  nonrelativistic version of the heuristic  procedure used in Ref. \cite{chew81}, which allows to obtain solutions for anisotropic matter from known solutions for isotropic matter; that is,
\begin{equation}
\Delta=Cf(r)\rho \,r^N,
\end{equation}
where $C$ is a parameter which measures the anisotropy; the function $f$ and
the number $N$ are to be specific for each model. Following that procedure the ansatz 
\begin{equation}
f r^{N-1}= \frac{d\Phi}{dr},
\end{equation}
leads to 
\begin{equation}
F=2C\Phi.
\end{equation}
Thus, Eq. (\ref{le1}) can be easily reduced to
\begin{equation}
\frac{d^2w}{dz^2}+\frac{2}{z}\frac{dw}{dz}+hw^n=0,
\label{le2}
\end{equation}
where $h=1-2C$.  For simplicity, we assume $h$ to be constant throughout the sphere, which of course does not imply the constancy of either pressure. {Observe that Eq.  (\ref{le2}) is the same as the Fowler equation \cite{f30}
when $\beta_0=-h$ and $\omega=0$ in the notation of \cite{bv91} [see Eq. (2.2) in this last reference]. However, it should be noticed that Eq. (\ref{le1}) is more general than the Fowler-Emden equation.}

Now, we proceed to integrate numerically with the boundary conditions
$$w(0)=1, \,\,\,\, \frac{dw}{dz}(0)=0,$$
with $n=3$, which represents a relativistic WD.

Figure 1 displays the dimensionless variable $w$ as a function of the dimensionless variable $z$ for different values of $h$.
Figure 2 shows the Chandrasekhar mass ratio (with respect to the isotropic mass $M_{Ch}$)
\begin{equation}
\frac{M}{M_{Ch}}=\frac{z_3^2\left(dw/dz\right)_{z_3}}{\,\,\,\,\left[z_3^2\left(dw/dz\right)_{z_3}\right]_{h=1}}
\end{equation}
as a function of the anisotropy parameter $h$.

Following the same ansatz for the heuristic model in the case $\gamma=1$, the  Lane-Emden equation becomes
\begin{equation}
\frac{d^2w}{dz^2}+\frac{2}{z}\frac{dw}{dz}=he^{-w},\label{le3}
\end{equation}
where
\begin{equation}
\Phi=\frac{K}{h}w,
\end{equation}
and from Eq. (\ref{2iso}) we get
\begin{equation}
\rho=\rho_c e^{-h\Phi/K}.
\end{equation}
Equation (\ref{le3}) has to be integrated
with the following central conditions: 
$$w(0)=0, \,\,\, \frac{dw}{dz}(0)=0.$$
Figure 3 displays the solution for different values of $h$.  As for polytropes with $n\ge 5$, the isothermal sphere consists of an ideal gas which has infinite radius. 

It is remarkable, although not general, to observe the invariance of scale of 
Eqs. (\ref{le2}) and (\ref{le3}). 
In fact, if we redefine the dimensionless spatial variable
\begin{equation}
\zeta=\sqrt{h} z, \label{resca}
\end{equation}
we recover the ``isotropic'' Lane-Emden equation, with all the special cases
or analytic solutions for $n=0,\,1,\, 5$. Doing so, it is easy to see where the surface is for any $h$ from calculation for $h=1$. This became obvious when we considered the asymptotic expansion near $z=0$ (for $\gamma\ne1$), rendering
\begin{equation}
w=1-\frac{1}{6}\zeta^2 + \frac{n}{120} \zeta^4 + \dots
\end{equation}
Thus, calculations in terms of $\zeta$ lead to an invariant $M_{Ch}$. But physically, the Chadrasekhar mass can clearly be stretched (or shrunk) with the anisotropy $h$. 
The same rescaling [Eq. (\ref{resca})] works for Eq. (\ref{le3}) with a different expansion near $\zeta=0$. 
\section{conclusions}
We have established the general framework foe modeling polytropes in the presence of anisotropic pressure. As mentioned in the Introduction, we undertook this task motivated by the conspicuous presence of such an anisotropy in compact objects and its influence on their structure.

We also obtained some specific models  based on a heuristic ansatz used many years ago to handle the anisotropy of pressure. The main purpose of that modeling was to bring out, in an explicit way, the influence of local anisotropy in such an important problem as the Chandrasekhar mass limit. We do not know if the inferred super-Chandrasekhar white dwarfs from collected data \cite{howell, scalzo10, scalzo12, ks11, hksn12, dm13}
are the result of anisotropy as considered here. But this
interesting matter and the physical conditions for real stars deserve more
attention elsewhere.

We have also shown (under the same ansatz) how the $\gamma=1$ case  is affected by anisotropy; this might be of interest in the discussion of the Sch\"onberg-Chandrasekhar limit.

Of course, for the modeling of specific astrophysical objects, full information about  the anisotropy ($\Delta$) has to be provided. 

{When the approach to anisotropy comes from kinetic theory by means of the Vlasov-Poisson system,  other equations such as Emden-Fowler can be derived from the generalized polytrope equation (see Ref. \cite{heinzletal} and references therein). All this is within the context of Newtonian gravity and spherical symmetry.}

Finally, we want to stress that all we have done here requires spherical symmetry, at least as an approximation. {It is possible that this symmetry can be broken by a strong magnetic field, rendering the distribution anisotropic and nonspherical \cite{22}. In such a case, of course, the method presented here does not apply.}

\section*{Acknowledgments}

{W. B. wishes to thank the Departamento de F\'\i sica Te\'orica e Historia de la Ciencia, Universidad del Pa\'\i s Vasco, for hospitality, especially J. 
Ib\'a\~nez and A. Di Prisco; and also the Intercambio Cient\'\i fico Program, U.L.A., for financial support.}

{\bibitem{andreasson} H. Andr\'easson, Living Rev. Relativity, {\bf 14}, 4 (2011).}

\bibitem{15} J. C. Kemp, J. B. Swedlund, J. D. Landstreet and  J. R. P. Angel, Astrophys. J., {\bf 161}, L77 (1970).

\bibitem{16} G. D. Schmidt and  P. S. Schmidt, Astrophys. J., {\bf 448}, 305 (1995).

\bibitem{17} A. Putney, Astrophys. J., {\bf 451}, L67 (1995).

\bibitem{18} D. Reimers, S. Jordan, D. Koester, N. Bade, Th. Kohler and L. Wisotzki, Astron. Astrophys., {\bf 311}, 572 (1996).

\bibitem{19} A. P. Martinez,   R. G. Felipe and  D. M. Paret, Int. J. Mod. Phys. D, {\bf 19}, 1511 (2010).

\bibitem{22} M. Bocquet, S. Bonazzola, E. Gourgoulhon and J. Novak, Astron. Astrophys., {\bf 301}, 757 (1995).

\bibitem{23} M. Chaichian,  S. S. Masood, C. Montonen, A. Perez Martinez and H. Perez Rojas, Phys. Rev. Lett, {\bf 84}, 5261 (2000).

\bibitem{24} A. Perez Martinez, H. Perez Rojas and H. J. Mosquera Cuesta,  Eur. Phys. J. C, {\bf 29}, 111 (2003).

\bibitem{25} A. Perez Martinez, H. Perez Rojas and H. J. Mosquera Cuesta, Int. J. Mod. Phys. D, {\bf 17}, 2107 (2008).

\bibitem{26} E. J. Ferrer, V. de la Incera, J. P. Keith, I. Portillo and P. L. Springsteen, Phys. Rev. C., {\bf 82}, 065802 (2010).

\bibitem{21} U. Das and B. Mukhopadhyay, Int. J. Mod. Phy. D, {\bf 21}, 1242001 (2012).

\bibitem{20} A. Kundu and  B. Mukhopadhyay, Mod. Phys. Lett A, {\bf 27}, 1250084 (2012).

\bibitem{27}  I. Suh and  G. J. Mathews, Astrophys. J., {\bf 530}, 949 (2000).

\bibitem{chew81} M. Cosenza, L. Herrera, M. Esculpi and L. Witten, J. Math. Phys., {\bf 22}, 118 (1981).

\bibitem{chew82} M. Cosenza, L. Herrera, M. Esculpi and L. Witten, Phys. Rev. D, {\bf 25}, 2527 (1982).

\bibitem{f30} R. H. Fowler, Mon. Not. R. Astron. Soc., {\bf 91}, 63 (1930).

\bibitem{bv91} O. P. Bhutani and K. Vijayakumar, J. Austral. Math. Soc. Ser. B {\bf 32}, 457 (1991).
 
{\bibitem{howell} D. A. Howell {\it et al.}, Nature (London), {\bf 443}, 308 (2006).}

{\bibitem{scalzo10} R. A. Scalzo {\it et al.},  Astrophys. J., {\bf 713}, 1073 (2010).}

{\bibitem{scalzo12} R. A. Scalzo {\it et al.},  Astrophys. J., {\it 757}, 12 (2012).}

{\bibitem{ks11} A. Kashi and N. Soker, Mon. Not. R. Astron. Soc., {\bf 417}, 1466 (2011).}

{\bibitem{hksn12} I. Hachisu, M. Kato, H. Saio and K. Nomoto, Astrophys. J., {\bf 744}, 69 (2012).}

{\bibitem{dm13} U. Das and B. Mukhopadhyay, Phy. Rev. Lett, {\bf 110}, 071102 (2013).}

{\bibitem{heinzletal} J. M. Heinzle,  A. D. Rendall, and C. Uggla, Math. Proc. Camb. Philos. Soc. {\bf 140}, 11 (2006).}

\end{document}